\documentstyle[draft,psfig]{mnusa}

\begin{document} 

\title[Photon Scattering in Schwarzschild Spacetimes]
{Photon Scattering by Relativistic Flows in Schwarzschild
Spacetimes. I.\ The Generation of Power-Law Spectra}
\author[H.\ Papathanassiou, D.\ Psaltis]
{Hara Papathanassiou${}^1$, Dimitrios Psaltis${}^{2,3}$\\
${}^1$International School for Advanced Studies (SISSA), Via Beirut 
2-4, 34013 Trieste, Italy\\ 
${}^2$Center for Space Research, Massachusetts Institute of Technology,
Cambridge, MA 02139, U.S.A.\\
${}^3$Harvard-Smithsonian Center for Astrophysics, 
60 Garden St., Cambridge, MA 02138, U.S.A.}

\date{\today}

\maketitle
\begin{abstract}

We study the spectra generated as a result of bulk Comptonization by
relativistic electrons in radial flows onto compact objects. We solve
numerically the general-relativistic radiative transfer equation in
the Schwarzschild spacetime, in steady-state and under minimal assumptions. We show that
power-law spectra result from multiple scatterings, in a way similar
to thermal Comptonization. We also find that photon-electron
interactions taking place near the black-hole event horizon affect
very little the emerging spectra. We, therefore, argue that bulk
Comptonization spectra do not carry distinguishing signatures of the
compact object around which they are produced.  We examine the
dependence of the spectra on simplifications often employed regarding
the spacetime geometry, the distribution of photon sources, and the
boundary conditions.  We show that the existence of trapped
characteristics around a black hole reduces the efficiency of
Comptonization and that general relativistic effects identically
cancel bulk Comptonization effects for a free-falling flow and in the
limit of infinitesimal mean-free path.  As a result, we find that
neglecting the curved spacetime geometry leads to overestimating the
high-energy flux by up to an order of magnitude.  Finally, we
demonstrate that the spectrum from accretion onto a neutron star
depends sensitively on the imposed boundary conditions, while that
from a black hole is immune to such choices.
\end{abstract}


\section{INTRODUCTION} 
\label{sec:intro}

Compton scattering of photons in media with non-negligible bulk
velocities is thought to be a significant, or even the dominant,
radiative process in a number of astrophysical settings. Examples
include the early Universe (the kinematic Sunyaev-Zeldovich effect;
see, e.g., Rephaeli 1995), jets in blazars (see, e.g., Sikora,
Begelman \& Rees 1994), gamma-ray bursts (see, e.g., Lazzati et al.\
2000; Madau \& Thompson 2000), and accretion flows onto galactic
compact objects (see, e.g., Payne \& Blandford 1981). The interaction
of neutrinos with fast-moving electrons is also thought to determine
the dynamics and fate of supernova explosions and super-critical
accretion flows onto neutron stars (see, e.g., Fryer, Benz \& Herant
1996; Burrows et al.\ 2000).

Studying any of the above problems requires the solution of the
kinetic equation that describes scattering of massless particles by
fast-moving electrons, in which relativistic effects introduce very
strong angular and photon-energy dependences. Because of such
complexities, most previous studies have made a number of
approximations in treating this radiative transfer problem (see
Psaltis \& Lamb 1997 and references therein). The kinetic equation is
often expanded to different orders in the electron bulk velocity
$\beta$ (in units of the speed of light) and/or the photon energy
$\epsilon/m_{\rm e}$ (where $m_{\rm e}$ is the electron rest mass) and is
truncated keeping only terms up to a low order \cite{BP81,MK92,TMK97}. 
Furthermore, general
relativistic effects are most usually neglected (but see Schmid--Burgk 1978; Zane et al.\ 1996; Titarchuk \& Zannias 1998; Laurent \&
Titarchuk 1999). Such analyses show that the effects of the odd- and
even-order terms in $\beta$ are qualitatively and quantitatively
different and their relative importance is determined not only by the
magnitude of the bulk velocity but by the properties of the radiation
field and the flow as well (Psaltis \& Lamb 1997, 2000).

The configuration in which Compton scattering of photons by
relativistic electrons has been studied extensively in the past is
that of accretion flows onto compact objects. Recent motivation for
such studies comes from the observed power-law shapes and the absence
of any detectable cut-offs (up to energies comparable to the rest mass
of the electron) in the $\gamma$-ray spectra of galactic accreting
black holes (Grove et al.\ 1998). Attributing these spectra to
Comptonization by the relativistic electrons in the accretion flow
could provide a spectral signature for the existence of a black hole
in a galactic system and even lead to the measurement of its mass (as
suggested in, e.g., Chakrabarti \& Titarchuk 1996; Shrader \&
Titarchuk 1998, 1999; Laurent \& Titarchuk 1999; Borozdin et al.\
1999).

Various analytical \cite{PB81,MK92,TZZN96,TZ98,ZL00}
and numerical (Zane et al.\
1996; Titarchuk et al.\ 1997; Laurent \& Titarchuk 1999; Psaltis 2000)
treatments have been employed in previous studies of bulk
Comptonization. However, the results of several analytic solutions
have been recently questioned because of the inadequate order to which
the transfer equations was truncated (see discussion in Psaltis \&
Lamb 1997, 2000) or of the inappropriate mathematical methods employed
(see discussion in Zampieri \& Lamb 2000).  Moreover, previous
numerical studies have either considered a limited number of cases (as
in, e.g., Zane et al.\ 1996) or used approximate treatments which
severely affected the properties of the solutions (see discussion in
Psaltis \& Lamb 2000).

With this paper we initiate a study of Compton scattering in
relativistic accretion flows and winds, performed under a minimal set
of assumptions and approximations, in order to address a number of
questions that previous treatments raised. We solve numerically the
kinetic equation for massless particles in a Schwarzschild spacetime
derived by Lindquist (1966), using an iterative integration along the
curved photon characteristics \cite{S-B78,Zane96}.
In this first study, we investigate the effects on the emerging
spectra of properties related  to the flow velocity, the metric,
the boundary conditions, and the distribution of sources in the
flow. We only assume that the systematic down-scattering of photons as
well as the Klein--Nishina corrections to the scattering cross section
(which are of the same order) are negligible. Therefore, we cannot
address issues related to the high-energy cut-offs of the spectra and
their shapes at photon energies comparable to, or higher, than the
electron rest mass; we will discuss these elsewhere.

\section{RADIATIVE TRANSFER IN A SCHWARZSCHILD SPACETIME} 
\label{sec:problem}

Throughout this paper we use geometric units ($c = G =1$) and describe
the radiation field in terms of the photon occupation number
$f(r,\mu,\epsilon)$. We assume that the spacetime is static and
spherically symmetric and, therefore, the occupation number depends
only on the radius $r$ (hereafter normalised to $2 M$, where $M$ is
the mass of the central object), the cosine $\mu$ of the angle between
the radial direction and the photon propagation vector, and the photon
energy $\epsilon$ appropriately normalised and measured in the
local rest frame. We define the necessary moments of the photon field
\begin{eqnarray} 
J(r,\epsilon) & \equiv & \frac{1}{2}\int_{-1}^1 f(r,\mu,\epsilon) 
\epsilon^3 d\mu\;, 
\label{eq:J_v}\\ 
H(r,\epsilon) & \equiv & \frac{1}{2} \int_{-1}^1 f(r,\mu,\epsilon)  
\epsilon^3 \mu d\mu\;, 
\label{eq:H_v}\\ 
K(r,\epsilon) & \equiv & \frac{1}{2} \int_{-1}^1 f(r,\mu,\epsilon)  
\epsilon^3 \mu^2 d\mu\;, 
\label{eq:K_v} 
\end{eqnarray} 
so that the energy flux emerging from the flow, which will be the 
primary quantity of interest, is 
\begin{equation} 
F(\epsilon)=4\pi\lim_{r\rightarrow\infty}H(r,\epsilon)\;. 
\label{eq:FF_v} 
\end{equation} 

In what follows, we assume $\epsilon\ll m_{\rm e}$ and
neglect any polarisation-dependent effects.  The steady-state photon
kinetic equation for the photon occupation number in a spherically
symmetric spacetime is (Lindquist 1966)
\begin{eqnarray} 
\displaystyle 
\sqrt{- g_{oo}} \gamma (\mu +\beta) \frac{\partial{f}}{\partial{r}} &+& 
\nonumber\\
 \sqrt{- g_{oo}} \gamma (1-\mu^2)   
\left[\frac{1 + \beta \mu}{r} +g_{oo} \gamma^2 (\mu +\beta)  
\frac{\partial{\beta}}{\partial{r}}\right] 
\frac{\partial{f}}{\partial{\mu}} & - & 
\nonumber\\ 
\sqrt{-g_{oo}} \gamma  
\left[\frac{\beta (1-\mu^2)}{r} - g_{oo} \gamma^2 \mu (\mu +\beta)  
\frac{\partial{\beta}}{\partial{r}} \right] \epsilon \frac{\partial{f}} 
{\partial{\epsilon}} & = & 
\nonumber \\ 
\qquad\qquad\qquad  
\frac{\eta(r,\epsilon)}{\epsilon^3} -\chi(r,\epsilon) f\;, 
\label{eq:rel_ss} 
\end{eqnarray} 
where $\gamma\equiv(1-\beta^2)^{-1/2}$ is the Lorenz factor,
$g_{oo}=-(1-1/r)$ for Schwarzschild geometry, and $g_{oo} = -1$ for flat
geometry. 
In the limit $\epsilon \ll m_{\rm e}$,
we include the effects of scattering by setting
\begin{equation}
\chi(r,\epsilon)=\chi_{\rm a}(r,\epsilon)+n_{\rm e}(r)\sigma_{\rm T}
\label{eq:chi}
\end{equation}
and
\begin{eqnarray} 
\eta(r,\epsilon)&=&\eta_{\rm e}(r,\epsilon)
+\frac{3}{8} n_{\rm e}\sigma_{\rm T} 
\left[\left(3-\mu^2\right)J(r,\epsilon)\right.\nonumber\\
& & \qquad\qquad+ 
\left.\left(3\mu^2-1\right)K(r,\epsilon)\right]\;, 
\label{eq:eta}
\end{eqnarray}
where $\eta_{\rm e}(r,\epsilon)$ and $\chi_{\rm a}(r,\epsilon)$ are
the emission and absorption coefficients, $\sigma_{\rm T}$ is the
Thomson scattering cross section, and $n_{\rm e}(r)$ is the electron
number density, all evaluated in the local rest frame. By writing $f$
as a full differential of the path length $s$ along each photon ray,
Eq.~[\ref{eq:rel_ss}] simplifies to (see Schmid--Burgk 1978 and Zane
et al.\ 1996 for the details of the method which we summarise here for
completeness)
\begin{equation} 
\frac{d f}{d s} = \frac{\eta}{\epsilon^3} - \chi f\;,  
\label{eq:df_ds} 
\end{equation}  
where  
\begin{eqnarray}
\frac{d r}{d s} &\hskip-2mm =  & \hskip-2mm\sqrt{-g_{oo}} \gamma (\mu +\beta) 
\label{eq:dr_ds}\\ 
\frac{d \mu}{d s} & \hskip-2mm =  & \hskip-2mm\sqrt{-g_{oo}} \gamma  (1 -\mu^2)  
\left[ \frac{1 +\beta \mu}{r}  
+ g_{oo} \gamma^2 (\mu +\beta) \frac{\partial{\beta}}{\partial{r}}\right] 
\label{eq:dm_ds}\\ 
\frac{d \epsilon}{d s} &\hskip-2mm = &\hskip-2mm \sqrt{-g_{oo}} \gamma  
\left[ \frac{\beta (1 -\mu^2)}{r}  
- g_{oo} \gamma^2 \mu (\mu +\beta)  
\frac{\partial{\beta}}{\partial{r}}\right] \epsilon\;.\; 
\label{eq:de_ds} 
\end{eqnarray} 

Algebraic manipulation of 
Eqs.~[\ref{eq:df_ds}]--[\ref{eq:de_ds}] results in the simple 
ordinary differential equation (Zane et al.\ 1996) 
\begin{equation} 
\frac{d f}{d r} = \frac{\eta/\epsilon^3 - \chi f} 
{\sqrt{-g_{oo}} \gamma (\mu +\beta)}\;, 
\label{eq:df_dr} 
\end{equation} 
where  
\begin{equation} 
\mu = \frac{ g_{oo} b^2 \beta \gamma^2 \pm r \sqrt{r^2 +g_{oo} b^2}} 
{r^2 - g_{oo} b ^2 \beta^2 \gamma^2 }\;, 
\label{eq:mu_char} 
\end{equation} 
\begin{equation} 
\epsilon = \frac{\epsilon_{\infty}}{\sqrt{-g_{oo}} \gamma (1 +\beta \mu)}\;.
\label{eq:e_char} 
\end{equation} 
The two last expressions make use of the two quantities that are conserved 
along a characteristic;  the impact parameter $b$, and the photon frequency 
as measured at infinity $\epsilon_{\infty}$. 
Characteristics with $b \le \sqrt{r_{\rm
 in}/|g_{oo}|}$ connect the inner boundary $r_{\rm in}$ with infinity.  In the
 Schwarzschild geometry, for every impact parameter with $b > 3 \sqrt{3}
 /2$, there exists one characteristic curve that reaches infinity and one that
 is trapped in the region $r < 1.5$. The trajectory with $b = 3 \sqrt{3}/2$
has a critical saddle point at radius $r =1.5$ and is a circle with this
radius.

We use different boundary conditions depending on the type of the
 characteristics. For the ones that reach infinity, we integrate from
 the outer boundary along an incoming ray ($\mu < 0$, using the
 negative-sign branch of solution~[\ref{eq:mu_char}]), setting
 appropriate boundary conditions for no external illumination, i.e.,
 $f(r_{\rm out},\mu<0,\epsilon)=0$.  For non-trapped characteristics
 that do not intersect the central object, the turning points are
 reached when $\mu=-\beta$ and the solution for $\mu$ simply switches
 to the positive-sign branch of Eq.~[\ref{eq:mu_char}]. For
 characteristics that reach infinity but do intersect the central
 object, we impose an additional boundary condition at $r_{\rm in}$,
 which depends on the problem under consideration. Typically, for the
 case of a neutron star, we set $f(r_{\rm
 in},\mu>0,\epsilon)=[1-\exp(-\epsilon/T_{\rm b})]^{-1}$, i.e., a
 blackbody of temperature $T_{\rm b}$, whereas for a black hole, we set
 $f(r_{\rm in},\mu>0,\epsilon)=0$, i.e., no illumination. Finally, we use
 similar boundary conditions for the trapped characteristics.
When including photon sources in the flow, the emissivity $\eta$ 
needs to be specified in Eq.~[\ref{eq:df_dr}], in addition to the
terms that describe scattering. In the idealised problems considered
here, we use two different photon sources that mimic more realistic
models of accretion flows. First, in order to model the effects of
volume emission, we adopt a blackbody source of photons of constant
temperature and an emission measure that is proportional to a power of
the electron density. Second, in order to model the effects of a
soft-photon input from an underlying accretion disk, we adopt a
blackbody source of photons with an emission measure proportional to
the viscous dissipation rate in an accretion disk, such that the
blackbody temperature at each radius is given by the standard
thin-disk solution (Shakura \& Sunyaev 1973)
\begin{equation}
T(r) = \left\{\begin{array}{lll}
T_{\rm in}\left(\displaystyle \frac{3}{r}\right)^{3/4}
\left(1-\sqrt{\displaystyle \frac{3}{r}}\right)^{1/4}\;, &\mbox{\rm for} &  r\ge 3\\
0\;, & \mbox{\rm for} & r<3,
\end{array}
\right.
\label{eq:diskemiss}
\end{equation}
where $T_{\rm in}$ is the temperature at $r=3$, inside which the
 accretion disk is assumed to be an inefficient source of photons.
  Note here that the photon-energy scale in our solutions is specified
 only by the boundary conditions or the photon sources, since we have
 assumed $\epsilon\ll m_{\rm e}$. As a result, in all
 calculations the photon energy is normalised to the blackbody
 temperature of the illuminating boundary (for some neutron-star
 cases), the temperature of the photon sources within the flow, or
 $T_{\rm in}$ (when a disk-like emissivity is assumed).

Finally, we typically set the electron velocity and density profiles
in the flow to their free-fall values, i.e., $\beta=-r^{-1/2}$ and
$n_e\sim r^{-3/2}$, and specify the normalisation of the latter
through the quantity
\begin{equation} 
\tau(r)\equiv \int_{r}^{r_{\rm out}} n_e(r) \sigma_{\rm T} dr\; 
\label{eq:tau_r}
\end{equation} 
evaluated at $r_{\rm in}$ which we call the optical depth. This is 
related to other quantities often used in describing such flows 
(Psaltis \& Lamb 1997). One such example is 
\begin{equation}
t_{\rm V}(r)\equiv -3 n_e(r)\sigma_{\rm T} \beta(r)  r 
= \frac{3}{2} \frac{\tau(r)}{\sqrt{r}}\;,
\label{eq:tv} 
\end{equation} 
which is equal to unity at the so-called photon trapping radius\footnote{This is the radius below which photons are advected inwards by the flow more efficiently than they are diffused outwards by scattering. It is not associated with the existence of trapped characteristics which are defined in \S\ref{sec:numerics}.}.
Another is 
\begin{equation} 
\frac{\dot{M}}{\dot{M}_{\rm E}} = \left(\frac{L}{\dot{M}}\right)
   r_{\rm in}^{1/2} \tau(r_{\rm in})\;, 
\label{eq:m_dot}
\end{equation} 
the mass accretion rate measured at infinity, in units of the
Eddington critical accretion rate (at which the outward radiation
force balances gravity) of an accretion flow that produces luminosity
$L$ with an efficiency $L/\dot{M}$.  Note, that when the accretion
rate becomes comparable to the Eddington critical rate, the assumption
of a free-fall velocity and density profiles is not justified, as the
outward radiation force on the accreting gas becomes non-negligible.

\section{NUMERICAL METHOD}
\label{sec:numerics}

We integrate Eq.~[\ref{eq:df_dr}] for a given impact parameter
and photon energy (as measured by an observer at infinity) by a 4-th
order Runge--Kutta method.  We use, in most cases, a logarithmically
equidistant grid in radius $r$ between the inner boundary $r_{\rm
in}$, which depends on the particular problem, and the outer boundary
$r_{\rm out}$ representing infinity. We fix the value of $r_{\rm out}$
to 30. For the case of a neutron star we set $r_{\rm in}=2.5$, while
for that of a black hole $r_{\rm in}=1.01$, to avoid the coordinate singularity
at the event horizon. When modelling a scattering problem of high optical
depth around a black hole, we use, instead, a grid that is
logarithmically equidistant in $r-1$.  This choice of the $r$ grid
allows us to reach arbitrarily close to the black hole event horizon and
with very fine resolution. This is necessary in order to deal with the
very steep derivatives present near the event horizon. Our grid in the
impact parameter $b$ is built with two different spacing prescriptions
(Zane et al.\ 1996).  For $ 0 \le b\le
\sqrt{r_*/|g_{oo}(r_*)|}$, where $r_* \equiv \rm{max}\left\{r_{\rm in},
1.5\right\}$, we use 
\begin{equation} 
b_j = \frac{r_*}{\gamma(r_*) \sqrt{-g_{oo}(r_*)}} \frac{\sqrt{1-\mu_j^2}}
{1 +\beta(r_*) \mu_j}\;,
\label{eq:b_anchored}
\end{equation}
with
\begin{equation}
\mu_j = -1 +(j -1) \frac{2}{N-1}
\label{eq:mu_anchored}
\end{equation} 
for the $j-$th point, which guarantees $N$ (usually 20) points spaced
 equally in $\mu$ at $r_*$\footnote{Choosing $r_*=1.5$ instead of $r_{\rm
 in}$, for the black-hole case, provides better angular sampling around $r=1.5$
 which would otherwise create an artificial spiky feature in the radial
 profiles of the various radiation quantities.}.  
  For $\sqrt{r_*/|g_{oo}(r_*)|} < b \le r_{\rm out}$, the grid on $b$ is
 constructed according to the method of tangent rays, i.e., so that the
 turning point ($r_{\rm min}= \sqrt{-b^2 g_{oo}}$) of each
 characteristic lies very close to an $r$ grid point.

For the photon energy grid we use an equidistant logarithmic grid that
covers the range of the comoving energies we want to calculate the
radiative processes in.  We fix the range from $\epsilon_{\rm min}=0.1$ 
to $\epsilon_{\rm max}=500$ (in units of the blackbody
temperature of the photon source or the illuminating inner
boundary). The gravitational and Doppler shifts experienced by the
photons as they propagate through the flow will cause radiation that
is produced locally in the above spectral range to appear at the
observer in a much wider range; of course, this range depends on the
compactness of the central object and is much wider for a black hole
than for a neutron star.  For example, if we place the inner boundary
at $r_{\rm in} = 1 +\delta r$, with $\delta r \ll 1$, photons with locally measured energy
$\epsilon$ in the [$\epsilon_{\rm min}$, $\epsilon_{\rm max}$]
interval will appear at the observer with an energy $\epsilon_\infty$
in the range
\begin{equation} 
\epsilon_{\infty,{\rm min}} =\frac{\delta r}{2}
\epsilon_{\rm min} \le \epsilon_{\infty} \le 2 \epsilon_{\rm max}
=\epsilon_{\infty,{\rm max}} 
\label{eq:freq_range} 
\end{equation} 
(see Eq.~[\ref{eq:e_char}]). For this reason, we maintain a
 logarithmic grid for $\epsilon_{\rm min} \le \epsilon \le 2
 \epsilon_{\rm max}$ and add a fixed number of logarithmically
 (but more sparsely) spaced points to cover energies below
 $\epsilon_{\rm min}$. 

As in all scattering problems, Eq.~[\ref{eq:df_dr}] is an
integro-differential equation and can, therefore, be most easily solved
by an iterative procedure. In all our calculations we neglect true
absorption and confine our attention to configurations with scattering
optical depths of order unity. It is, therefore, adequate to use a
simple variant of the $\Lambda-$iteration method (Mihalas 1978), which
typically converges after a few iterations. We have validated the
implementation of the numerical algorithm by comparing our solutions
to simple analytical results derived in appropriate limits and to the
numerical results reported for flat spacetimes and small
electron velocities in Psaltis (2000).

\section{RESULTS FOR NEUTRON STARS} 
\label{sec:NS}
We first model the transport of photons in an accretion flow that 
contains no photon sources but is illuminated from its inner boundary. 
 This configuration allows us to study the effects on the spectrum of 
the propagation of photons through a relativistic medium, independent 
of the potential complication introduced by the radial profiles of 
sources (cf. Psaltis \& Lamb 2000). Although idealised, our 
calculations are applicable to cold radial accretion flows onto 
neutron stars, if most of the accretion luminosity is released at the 
impact of the flow with the stellar surface. 

\begin{figure}
\centerline{ 
\psfig{file=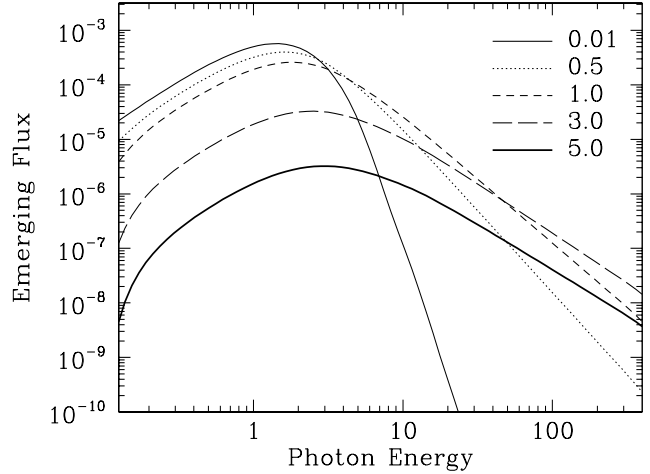,angle=-90,height=6.5truecm,width=8.5truecm}}
\caption{
Radiation spectra emerging from radial accretion flows onto a neutron 
star for representative optical depths. The flows are illuminated at their 
inner boundary by the same  black-body spectrum. The emerging flux is
in arbitrary units and the photon energy is in units of the
temperature of the illuminating blackbody spectrum.}
\label{fig:NS_spectra}
\end{figure} 

Fig.~\ref{fig:NS_spectra} shows the spectra emerging from such a flow
for different optical depths.  The results of the calculation
presented in this figure provide additional support to the suggestion
that the generation of a power-law high energy spectrum is a general
feature of Comptonization by relativistic inflows (see, e.g., Payne \&
Blandford 1981; note, however, that all early papers neglected the
terms that describe the effect we discuss here).
 
The physical reason behind the generation of the power-law spectral
tails is not readily obvious. For example, in the limit of very small
optical depth, very few photons interact with electrons more than once
and, therefore, it might appear reasonable to use a single-scattering
approximation in calculating the emerging spectrum.  If this were the
case, different photons interacting with electrons of different
velocities would gain different amounts of energy and, since the
electrons in the flow have a power-law velocity profile, they would
potentially generate a power-law spectral tail.  The result would then
be equivalent to the generation of a power-law spectrum produced in
optically thin, non-thermal plasmas, when low-energy photons are
scattered once by electrons with a power-law velocity distribution
(Coppi 1999).  On the other hand, the distribution of photon escape
times in the flow has an exponential tail at late times and this,
convolved with an exponential increase of photon energy per
scattering, can also produce a power-law high-energy tail as a result
of multiple scatterings (as suggested by the Monte-Carlo simulations
of Laurent \& Titarchuk (1999)). This would be similar to the generation
of power-law spectral tails in media that are optically thick to
(either thermal or non-thermal) Comptonization (Sunyaev \& Titarchuk
1980; Coppi 1999).

\begin{figure}
\centerline{
\psfig{file=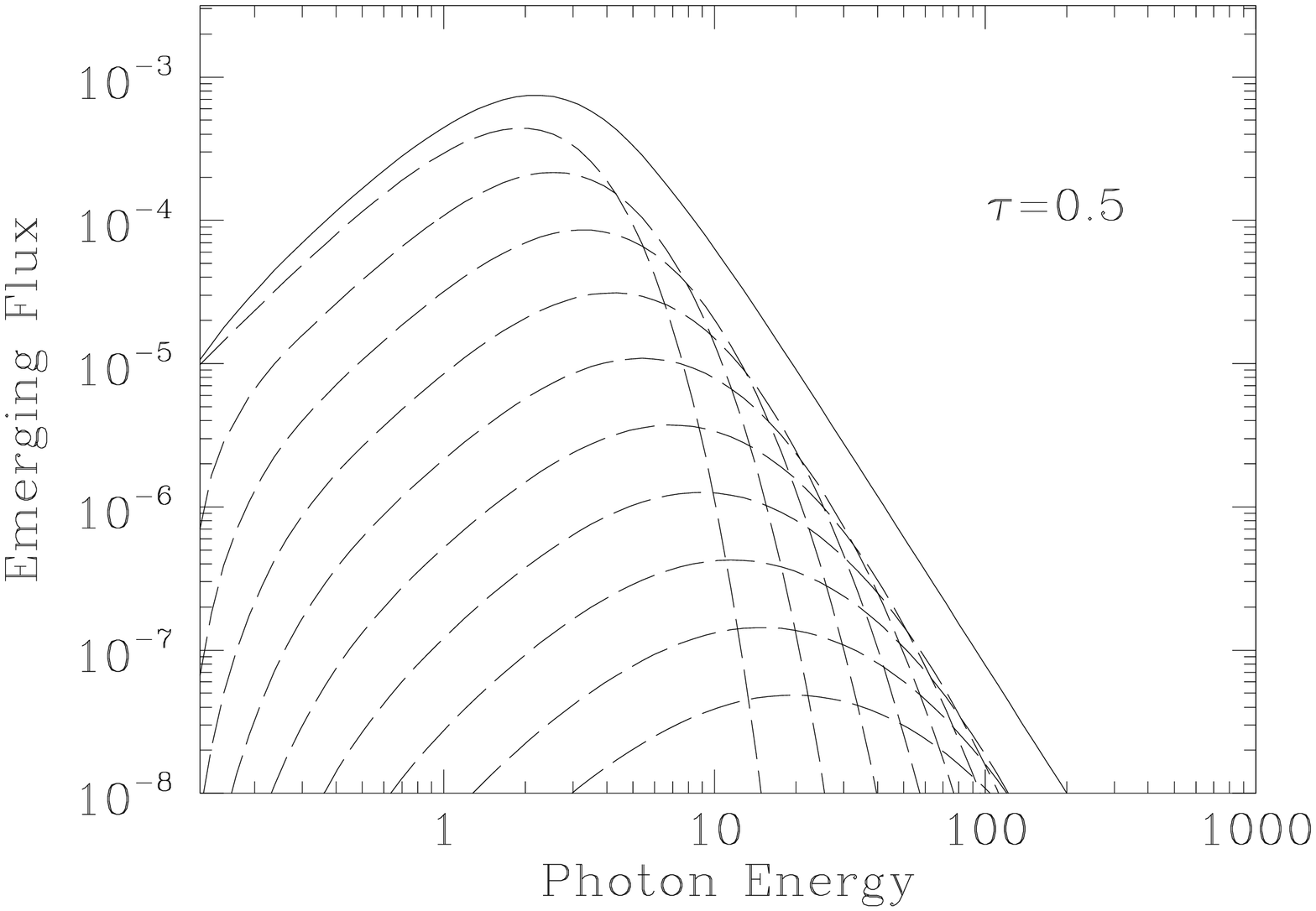,angle=-90,height=6.5truecm,width=8.5truecm}}
\centerline{
\psfig{file=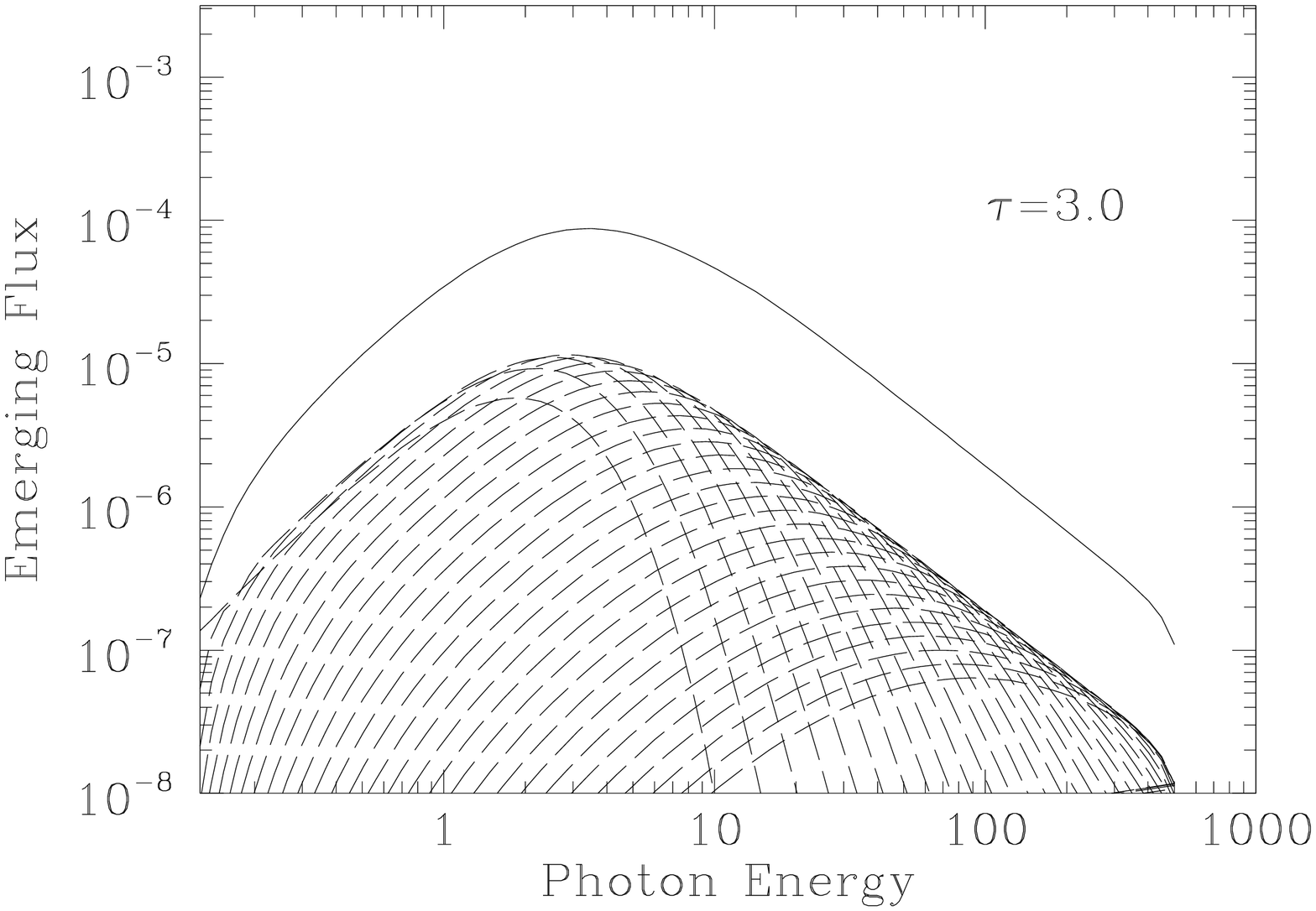,angle=-90,height=6.5truecm,width=8.5truecm}}
\caption{Decomposition into scattering orders (dashed lines) of the
energy spectra (solid lines) emerging from two of the accretion flows
presented in Fig.~1. The leftmost dashed line corresponds to the
zeroth scattering order, i.e., the photons that reached the observer
without having interacted with electrons. Each subsequent dashed line
to its right corresponds to an increasingly higher scattering order.}
\label{fig:scat_orders}
\end{figure} 

We can address the aforementioned question by calculating separately the
spectra of photons that emerge from a flow after having interacted
with electrons a given number of times. For this purpose, we decompose
the radiative transfer equation (\ref{eq:df_dr}) into a system of
equations for the individual scattering orders as
\begin{equation} 
\sqrt{-g_{oo}}\gamma(\mu+\beta)\frac{df_n}{dr}= 
\frac{\eta_{ n-1}}{\epsilon^3}-\chi f_{n}\;, 
\label{eq:dfn_dr} 
\end{equation} 
where $n = 1,2,$... represents the scattering order. We solve
Eq.~[\ref{eq:dfn_dr}] for the zeroth order setting $\eta_{-1}=0$ and
using the same boundary conditions as in the full problem, i.e.,
illumination from the inner boundary. We then solve the same equation
for each successive order $n$, calculating $\eta_{n-1}$ from the
solution of the previous order (through Eq.~[\ref{eq:eta}],) and
setting the appropriate boundary conditions for a medium that is not
illuminated from either boundary.

\begin{figure} 
\centerline{
\psfig{file=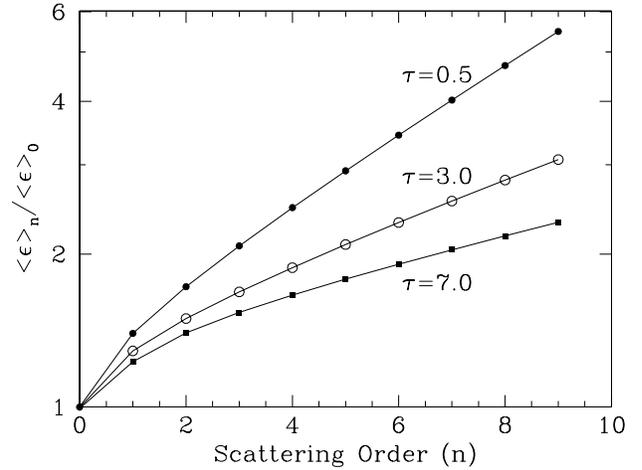,angle=-90,height=6.5truecm,width=8.5truecm}}
\caption{The average energy of photons emerging from the flows
of Fig.~1 after a {\em given\/} number of
scatterings. Results are shown for three representative optical depths.
The photon energies are normalised to the average
photon energy of photons that did not interact with the flow (which
corresponds to the zeroth scattering order).}
\label{fig:NS_av_energy}
\end{figure}

Fig.~\ref{fig:scat_orders} shows the decomposition into scattering
 orders of the energy spectra emerging from two flows with different
 optical depths. In both cases, the spectrum calculated for each
 scattering order is displaced and broadened compared to the spectrum
 of the previous order while the fractional change of the mean photon
 energy is comparable to an average value of $\beta^2$ in the flow.
This is demonstrated in Fig.~\ref{fig:NS_av_energy}, where the mean 
energy of photons emerging after each successive scattering order,
defined as
\begin{equation}
\langle \epsilon\rangle_{n} =
\frac{\int_0^\infty \epsilon^3 d\epsilon \int_{-1}^1 \mu d\mu  
\; f_{n}(r\rightarrow \infty,\epsilon,\mu)}
{\int_0^\infty \epsilon^2 d\epsilon \int_{-1}^1 \mu d\mu  
\; f_{n}(r\rightarrow \infty,\epsilon,\mu)}\;,
\label{eq:ave_energy} 
\end{equation} 
is plotted as a function of the scattering order $n$. After an initial
small number of scatterings, $\langle\epsilon\rangle_{n}$
increases exponentially with scattering order (i.e., producing a
straight line in Fig.~\ref{fig:NS_av_energy}), implying that the
average fractional energy change per scattering, $\delta
\epsilon/\epsilon$, remains constant and hence
$\langle\epsilon\rangle_{n}\sim e^{(\delta\epsilon/\epsilon)n}$.
The distribution, $f_{n} dn$, of the number of scatterings each
photon experiences in a spherical flow depends on the details of the
flow but often has an exponential tail, i.e., $f_{n}\sim
e^{-\alpha n}$ for $n\gg 1$ (see, e.g., Sunyaev \& Titarchuk 1980). As
a result, the convolution of these two exponentials gives rise to the
hard power-law spectrum of the form $f(\epsilon)\sim
\epsilon^{-\alpha(\epsilon/\delta\epsilon)-1}$, as seen in Fig.~1. It
is, therefore, the effect of multiple scatterings that produces the
power-law spectra at high photon energies, for both low and high
optical depths.

Note here that the average photon energy increases slower with
scattering order for the flows with the larger total optical depth
(see Fig.~\ref{fig:NS_av_energy}), a result that might appear at first
counterintuitive.  However, it can be understood as follows: the
quantity $\langle \epsilon \rangle_{n}$ measures the average
energy of photons that escape to infinity having experienced {\em
only\/} $n$ number of scatterings and not the average energy of all
photons after each successive scattering. Indeed, when the total
optical depth of the flow is low, photons that experience their $n-$th
scattering very close to the inner boundary of the flow, and hence
gain a lot of energy from the fast-moving electrons there, have a high
chance of escaping to infinity and, consequently, contributing to $\langle
\epsilon\rangle_n$. On the other hand, when the total optical depth of
the flow is high, only photons that experience their $n-$th scattering
away from the inner boundary of the flow, and hence gain a moderate
amount of energy from scattering off slower electrons, escape to
infinity and contribute to $\langle \epsilon\rangle_{n}$.

The emerging spectra at photon energies significantly larger than the
injection energy correspond to photons that have been scattered by
electrons a very large number of times and have therefore lost memory
of their initial energy and angular distribution. For this reason, the
slopes of the power-law tails depend only very weakly on the initial
energy of the photons or the reference frame in which the boundary
conditions are imposed.  These examples (together with the discussion
in Psaltis \& Lamb 1997) demonstrate the similarity between the
generation of power-law spectra tails in the flows considered here and
in hot but static scattering media (Sunyaev \& Titarchuk 1980).

\begin{figure} 
\centerline{ 
\psfig{file=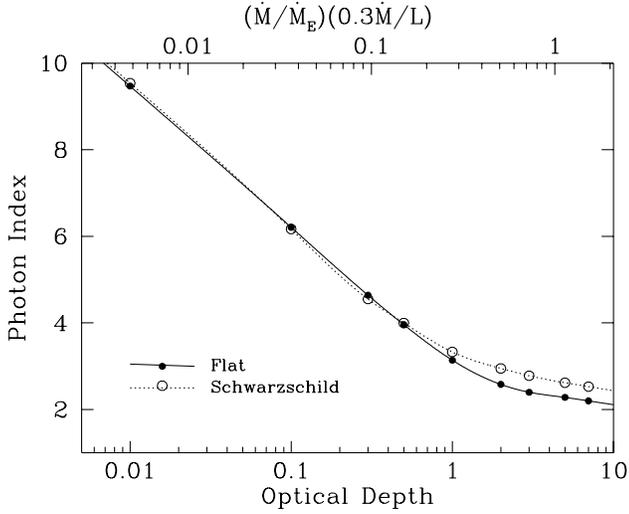,angle=-90,height=7truecm,width=8.5truecm}} 
\caption{
The optical depth dependence of the indices of the power-law tails
generated by the flows shown in  Fig.~1, as well
as for the same flows but calculated for a flat spacetime. Note that
the indices were calculated for the photon number spectra and not
 the energy spectra plotted in Fig.~1.}
\label{fig:NS_spec_index_metric}
\end{figure}

Fig.~\ref{fig:NS_spec_index_metric} shows the optical depth dependence 
of the photon index, defined as the slope of the
high-energy tail of the function $F(\epsilon)/\epsilon$ representing the photon number density.  As was
already obvious in Fig.~\ref{fig:NS_spectra}, the power-law tails
become flatter as the optical depth increases.  Note here that the
apparent saturation of the photon index at high optical depths is an
artefact of our neglecting the systematic down-scattering of
photons and the Klein--Nishina form of the scattering cross section. 
Taking these effects into account (which are of order
$\epsilon/m_{\rm e}$ and higher) would produce, for high optical
 depths, a prominent Wien peak at energies comparable to the mean
 electron kinetic energy (Psaltis \& Lamb 2000) and therefore affect the
 power-law nature of the emerging spectra. Moreover, at the corresponding
 high inferred accretion rates, the radial profiles of the electron density
 and velocity will not be the free-fall profiles assumed here.

Fig.~\ref{fig:NS_spec_index_metric} also demonstrates that for a
free-falling medium and at high optical depths, the spectra calculated
for a Schwarzschild spacetime are steeper than the ones calculated for
a flat spacetime. This property can be traced back to the photon
kinetic equation (\ref{eq:df_dr}) and the effect on the emerging
spectra of the terms of different orders in the electron bulk velocity
$\beta$~(see also discussion in Zampieri \& Lamb 2000). Assuming a
general electron velocity $\beta$, denoting the free-fall velocity as
$\beta_{\rm ff}$, and expanding Eq.~[\ref{eq:df_dr}] up to second
order in velocity, we obtain
\begin{equation}
\left\{\mu+ \beta+ \frac{1}{2}(\beta^2-\beta_{\rm ff}^2)\mu + 
{\cal O}[\beta(\beta^2-\beta_{\rm ff}^2)]\right\} 
\frac{df}{dr}\hskip-1mm = \hskip-1mm \frac{\eta}{\epsilon^3}  - \chi f,\;
\label{eq:dfdr_s} 
\end{equation} 
for a Schwarzschild geometry, and 
\begin{equation} 
\left[\mu+\beta+ \frac{1}{2}\beta^2\mu+{\cal O}(\beta^3)\right] 
\frac{df}{dr} = \frac{\eta}{\epsilon^3} -\chi f\;, 
\label{eq:dfdr_f} 
\end{equation} 
for a flat geometry. 

The generation of a power-law tail in the emerging spectrum is
governed by the terms of order $\beta^2$ (Psaltis \& Lamb 1997, 2000),
which are the lowest order terms containing information about the
kinetic energy of the electrons that can be transfered to the photons.
Such terms appear both explicitly in
Eqs.~(\ref{eq:dfdr_s})--(\ref{eq:dfdr_f}) and implicitly through the
dependence of the characteristic curves, defined by
Eqs.~(\ref{eq:mu_char})--(\ref{eq:e_char}), on $\beta^2$; the latter
contribution is negligible when the photon mean-free path is
significantly smaller than any characteristic length-scale of the
system. For a free-falling atmosphere in a Schwarzschild geometry, the
explicit terms of order $\beta^2$ in the photon kinetic equation are
identically cancelled by the terms of order $\beta_{\rm ff}^2$ that
describe the gravitational redshift. Note, however, that the photons
escaping to infinity have experienced --by definition-- the last
scattering in regions of very large photon mean-free path and,
therefore, the cancellation of the systematic upscattering by general
relativistic effects is not severe. As a result, the efficiency of the
energy exchange between photons and fast moving electrons is reduced
when general relativistic effects are taken into account and the effect is
pronounced in flows with high optical depths, where a photon has suffered
many scatterings before escaping.

\begin{figure} 
\centerline{ 
\psfig{file=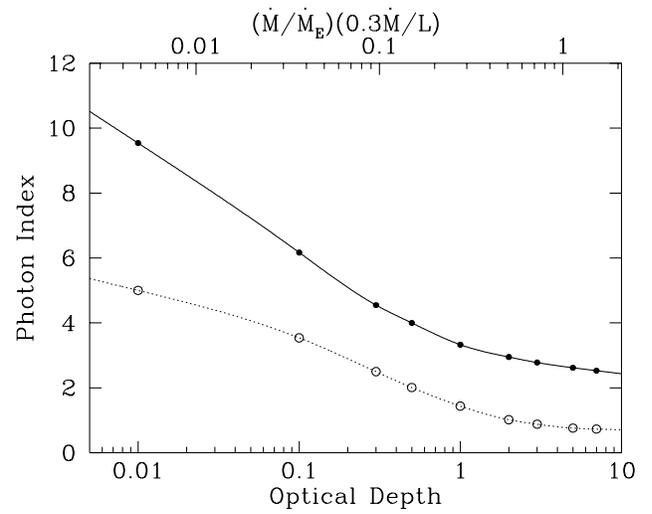,angle=-90,height=7truecm,width=8.5truecm}} 
\caption{
The optical depth dependence of the photon indices of the 
power-law tails generated by a free-falling medium onto a neutron star.
 The solid and dotted lines 
correspond to an absorptive and a semi-reflective inner boundary 
condition respectively, as described in the text.} 
\label{fig:NS_spec_index_IBC}
\end{figure} 

As discussed above, the calculated photon indices are rather
independent of the injection energy of the photons. However, they do
depend strongly on the choice of the inner boundary condition for the
flow. This is the case since the inner boundary condition
affects the distribution of photon escape times from the flow and,
hence, their average energy gain.  Fig.~\ref{fig:NS_spec_index_IBC}
compares the results presented earlier with the power-law slopes of the
spectra emerging from similar flows in which we have imposed a
semi-reflecting inner boundary condition, i.e., 
\begin{equation}
f(r_{\rm in},\mu>0,\epsilon)=f(r_{\rm in},-\mu,\epsilon)+ 
\frac{1}{\exp(\epsilon/T_{\rm b})-1}\;. 
\end{equation} 
The spectra emerging from the flows with semi-reflecting 
boundaries are significantly flatter because the photons can be 
reflected a number of times between the flow and the inner boundary, 
gaining more energy, before emerging from the flow. The effect is very
 large, as the emerging spectrum can be very flat (index$\simeq 0.5$)
 or steep (index$\simeq 2.5$) for the same (large) value of the optical depth
 but for different boundary conditions.

\section{RESULTS FOR BLACK HOLES} 
\label{sec:BH}

\begin{figure} 
\centerline{
\psfig{file=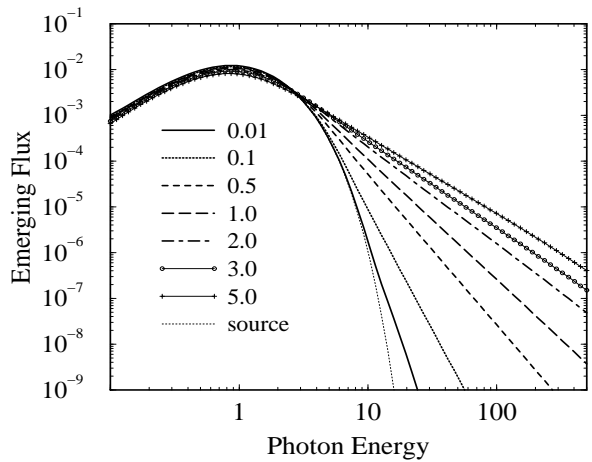,angle=0,height=6.5truecm,width=8.5truecm}} 
\caption{Radiation spectra emerging from scattering in radial accretion 
flows onto a black hole with different optical depths. The distribution
 of photon sources resembles the black-body emission from an underlying
accretion disk (Eq.~[15]). The emerging flux is in arbitrary units 
and the photon energy is in units of the temperature of the accretion 
disk at its inner boundary. The spectrum resulting in the absence of 
scattering (thin line) is also shown for comparison.}
\label{fig:BH_spectra_disk}
\end{figure}

In this section we model the transport of photons in a radial
accretion flow onto a black hole. In such a flow, the source of soft
photons depends on the detailed properties of the accretion
model and various simple expressions have been used so far in previous
studies.  Fig.~\ref{fig:BH_spectra_disk} shows the spectra
calculated for a disk-like source of photons
(Eq.~[\ref{eq:diskemiss}]), in a Schwarzschild spacetime, for
different values of the optical depth (Eq.~[\ref{eq:tau_r}]).  In order
to demonstrate the effect of Comptonization on the spectral shape, the
spectrum emerging from the flow in the absence of scattering has also
been included.  The resulting spectra consist of the soft source
spectrum smoothly extending to a hard power law tail which becomes
flatter with increasing optical depth, much like in flows illuminated
from the surface of the neutron star presented in \S4.

\begin{figure} 
\centerline{ 
\psfig{file=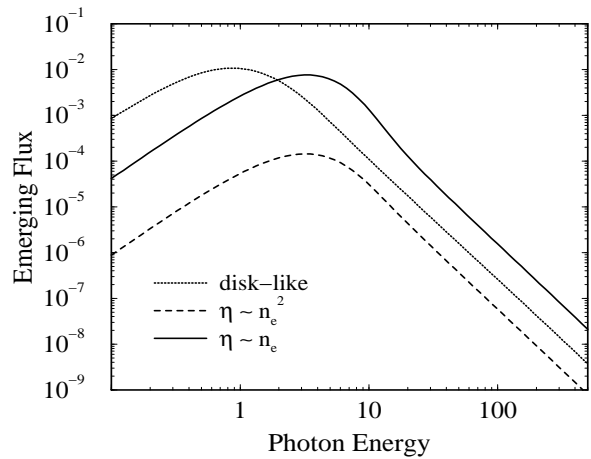,angle=0,height=6.5truecm,width=8.5truecm}}
\caption{Radiation spectra emerging from radial accretion flows onto
a black hole, with optical depth $\tau=1$, but different emissivities
($\eta$) of soft photons (discussed in the text). The index of the
power law at high energies is practically the same in all three
cases.}
\label{fig:BH_spectra_sources}
\end{figure}

The power-law indices of the calculated spectra depend very weakly, if at all,
 on the
radial distribution of the photon sources, as is evident in
Fig.~\ref{fig:BH_spectra_sources}.  This figure compares the spectra
emerging from flows of optical depth $\tau=1$ and for different
expressions for the photon source: a disk-like source of photons
(solid curve) and a volume emissivity that is proportional to the
first (dashed curve) and second (dotted curve) power of the electron
density. The close similarity of the power-law indices is yet another
demonstration of the fact that the spectral tails result from multiple
scatterings of photons by electrons which erase all memory of initial
photon distributions (see also discussion in \S\ref{sec:NS}).

Even though the basic mechanism for the generation of the power-law
 spectral tails is the same for  a flow around both a neutron star and a
 black hole, these systems differ in two respects: the location
 of the inner boundary and its radiative
 behaviour. The former brings out both the dramatic effect of the
 curvature of the spacetime and the extreme velocities, while the
 latter excludes the central object from any radiative contributions.

\begin{figure} 
\centerline{
\psfig{file=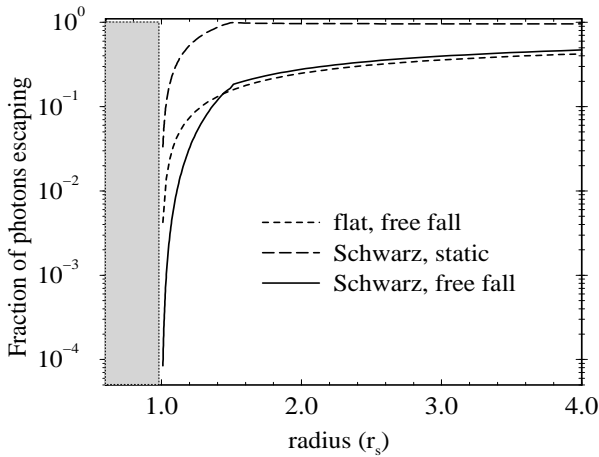,angle=0,height=6.5truecm,width=8.5truecm}}
\caption{The fraction of photons emitted by an isotropic source of
radiation at radius $r$ that escape to infinity, for combinations of different 
spacetime geometries and velocity fields.}
\label{fig:photon_escape}
\end{figure}

In an accretion flow around a black hole, the inner boundary lies at
the event horizon, in the vicinity of which the free-fall velocity
approaches the speed of light. Photons that propagate near the event
horizon are dragged inwards by the converging flow, even in the
presence of intense scattering that would tend to enhance their
diffusion outwards. Moreover, among photons with small impact
parameters (i.e., $b \le 3\sqrt{3}/2$), only those directed almost
radially outwards can escape the steep potential. The effects of the
velocity field and the spacetime geometry on the fraction of photons
that can escape to infinity from a distance $r$ from the centre of the
source are disentangled in Fig.~\ref{fig:photon_escape}.  For
illustration, given an isotropic source of photons located at radius
$r$, the fraction of escaping photons is calculated at the free
streaming limit, and corresponds to the fractional solid angle
subtended by the characteristics that reach radial infinity.  The
combination of the effects of velocity and geometry, in the
self-consistent picture of free falling material in a Schwarzschild
spacetime, causes a steep photon deficiency inwards of $r=1.5$, where
trapped photon trajectories exist.  From Fig.~\ref{fig:photon_escape},
it is evident that the properties of the flow and of the spacetime
inside $r =1.5$ affect very little the radiation that reaches the
observer. This would not be true if our description ignored the curved
geometry of the spacetime.

The other distinctive difference between a black hole and a neutron
star is the radiative behaviour of the inner boundary (which, in the case of
a neutron star, affects
strongly the emerging spectrum as discussed in \S4). The inner
boundary in the case of a black hole is its event horizon, out of which
no radiation can escape.  In order to isolate the effect of the
location of the inner boundary  
from that of the type of the inner boundary condition (i.e., the
radiative behaviour of the boundary), we performed a set of
calculations with all possible combinations of the two, some of 
which are unphysical.  

\begin{figure} 
\centerline{ 
\psfig{file=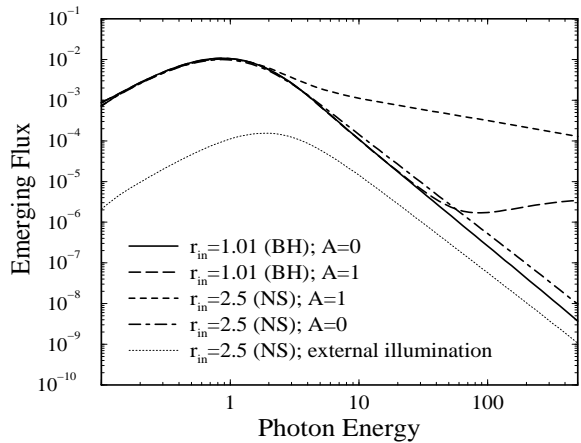,angle=0,height=6.5truecm,width=8.5truecm}}
\caption{Radiation spectra emerging from radial accretion flows with
the same optical depth $\tau=1$ but different locations and radiative
behaviours of the inner boundary. BH stands for a boundary appropriate
for a black hole, while NS stands for neutron star. $A=0$ ($A=1$)
corresponds to a fully absorptive (reflective) inner boundary.  The
case of external illumination with no photon sources within the flow
is also shown for comparison.}
\label{fig:ibc_effect}
\end{figure}

The results of the calculations, in which we vary the location and
behaviour of the inner boundary condition but keep the optical depth of
the scattering medium ($\tau=1$) and the functional form of the photon
sources (disk-like; see Eq.~\ref{eq:diskemiss}) fixed, are presented
in Fig.~\ref{fig:ibc_effect}. We use the albedo ($A$) to refer to
the radiative behaviour of the inner boundary condition.  The flattest
spectrum (short-dashed curve) corresponds to a configuration that is
realistic for a neutron star, with $r_{\rm in}=2.5$ and a totally
reflective inner boundary condition, i.e., $A = 1$.  The dot-dashed
curve shows the spectrum of a hypothetical neutron star that absorbs
all radiation that hits its surface without reemiting any of it
($A=0$). The range between these two curves is covered by similar
flows of intermediate albedoes.  The steeper spectrum in the latter
case is caused by the relative deficiency of photons available for
upscattering (see Fig.~\ref{fig:NS_spec_index_IBC} and its
discussion in \S\ref{sec:NS}). Accordingly, the spectrum resulting
from a similar flow onto a black hole (solid curve) is steep but not
considerably more so. The enhanced energy gain by the photons due to
the much higher velocities encountered closer to the event horizon
is compensated for by the shielding of the immediate neighbourhood of the
horizon by the combined effect of curvature and free-fall
velocity (demonstrated in Fig.~\ref{fig:photon_escape}).  The long-dashed
curve corresponds to a flow that reaches the  horizon with a reflective
inner boundary condition. This is certainly an unphysical
configuration; it is nevertheless instructive.  The resulting spectrum
coincides with that of the black hole with an absorptive inner
boundary condition, apart from a turnover at the hardest end which is,
however, totally artificial.  As we have already discussed in
\S\ref{sec:problem}, due to the coordinate singularity at the horizon,
the inner boundary is placed {\it very close to}, but not {\it at} the
horizon. The radiation that is reflected off this inner boundary
shows as the hard excess. A choice of an inner boundary closer to the
horizon (which can be achieved at a low computational cost with our
logarithmic grid in $r-1$) pushes this artificial excess to higher
photon energies. Apart from this artefact of the numerical method, it
is evident that the particular form of the inner boundary condition,
for the case of accretion onto a black hole, does not affect the emerging
spectrum. Indeed, the absorptive character of the inner boundary is
equivalent to the coincidence of this boundary with the black hole
event horizon, since the propagation of the photons in its vicinity is
determined entirely by the properties of the spacetime.

\begin{figure} 
\centerline{ 
\psfig{file=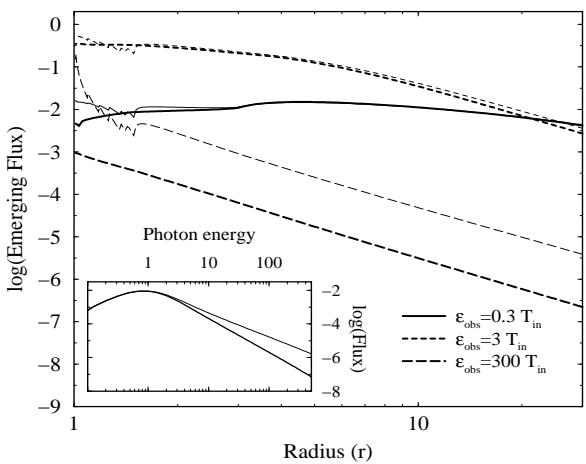,angle=-90,height=6.5truecm,width=8.5truecm}} 
\caption{Radial profiles of flux at three representative observed photon 
energies: one below the peak of the local emissivity (0.3~$T_{\rm in}$), one around its peak (3~$T_{\rm in}$), and one at the hard tail
(300~$T_{\rm in}$). Comparison between the results in a Schwarzschild (thick
curves) and flat (thin curves) spacetime.  The observed spectrum for
the two cases is shown in the insert.}
\label{fig:gr_flat_comparison}
\end{figure}

We turn now to examine the effect of neglecting the curvature of
the spacetime (as has been done in a large number of previous studies,
see discussion in \S\ref{sec:intro}) for the case of a flow that
reaches the event horizon of a black hole.  In
Fig.~\ref{fig:gr_flat_comparison}, we present the results of
calculations using the Schwarzschild metric (thick lines) or a flat
spacetime geometry (thin lines).  We performed both calculations for
disk-like photon sources (Eq.~[\ref{eq:diskemiss}]) and for an optical
depth of $\tau =3$.  The radial distribution of the flux is shown for
three representative observed photon energies: one around the peak of
the local emissivity function (3~$T_{\rm in}$ in this case), one below
the peak (0.3~$T_{\rm in}$) and one at the hard spectral tail
(300~$T_{\rm in}$). There is no substantial difference in the amount of soft
radiation reaching the observer between the calculations performed for the
two different spacetime geometries. This is true because the flux at a
certain radius and at frequencies around or below the spectral peak
mainly correspond to photons that  have not experienced any scattering with
electrons.  The hard spectral tail, though, which is the result of
multiple upscatterings, is significantly different between the two
calculations.  
Since no 
trapped photon characteristics exist in flat geometry (see \S\ref{sec:numerics}), a larger fraction of photons from 
radii close to the event horizon can escape and thus provide copious numbers 
of photons to be further upscattered.
This
effect compliments the one already discussed in \S4, i.e., that
general relativistic effects reduce the efficiency of bulk
Comptonization, identically cancelling it in the limit of very small
photon mean-free path. The net result is again a flatter power-law
spectrum for the calculation in a flat spacetime. This is
qualitatively the same as for the case of accretion onto a neutron
star (see \S\ref{sec:NS}) but more pronounced, as expected.  For
the cases shown in Fig.~\ref{fig:gr_flat_comparison}, the photon
index is $\simeq 2.45$ for the flat and $\simeq 3.03$ for
the Schwarzschild spacetime (compare with the difference of photon
spectral indices in the case of a neutron star shown in
Fig.~\ref{fig:NS_spec_index_metric}).

 Finally, Fig.~11 compares our results to previous calculations in
which the effects of general relativity were explicitly taken into
account. The photon index values we have obtained are consistent with the single case of
cold, radial accretion onto a black hole studied by Zane et al.\
(1996) and similar to the Monte Carlo solutions of Laurent \&
Titarchuk (1999) for low optical depths. At high optical depths, our
solutions correspond to slightly flatter spectra than those of Laurent
\& Titarchuk (1999) and the difference can be attributed to two
effects: First, we neglected the systematic down-scattering of photons
as well as the Klein--Nishina corrections to the scattering cross
section, both of which tend to reduce the efficiency of energy
exchange and hence produce steeper spectra. Second, the accretion
flows in the calculations of Laurent \& Titarchuk (1999) are truncated
at a radius smaller than in our calculations (i.e., at $r_{\rm
out}=3$) and therefore the energy gained by the interaction of photons
with electrons at larger radii is not taken into account. Finally, it
is unclear how to compare our results to the analytic solutions of
Titarchuk \& Zannias (1998), who calculated the eigenfunctions of the
transfer equation for flows with neither external illumination (Eq.~[24])
nor photon sources within the flow (Eq.~[16]).

\begin{figure} 
\centerline{
\psfig{file=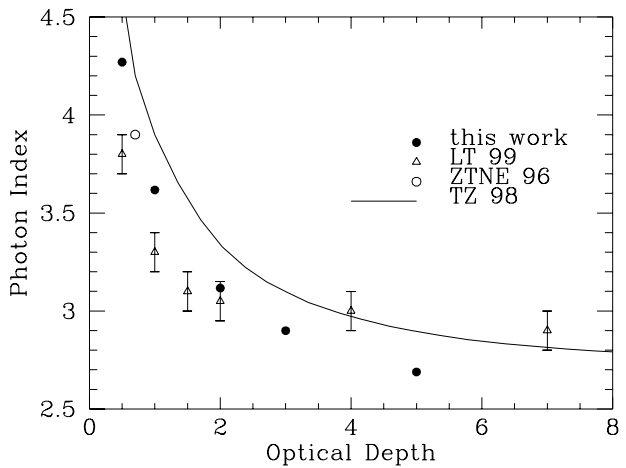,angle=-90,height=6.5truecm,width=8.5truecm}}
\caption{Comparison between different studies of the photon indices
calculated for bulk Comptonized spectra from spherically accreting
flows onto a black hole as a function of optical depth. The results of this work
correspond to the spectra shown in Fig~6. Comparisons are done with
the cold-flow solution of Zane et al.\ (1996; ZTNE 96), the analytic
calculation of Titarchuk \& Zannias (1998; TZ 98), and the
zero-temperature Monte-Carlo solutions of Laurent \& Titarchuk (1999;
LT 99). The points in the latter case correspond to accretion flows
with the same electron density profile as the current calculations
(note the difference in our definition of optical depth) and the error
bars represent the quoted 5\% uncertainty in the fitted power-law
indices.}
\label{fig:index_comparison}
\end{figure} 

\section{CONCLUSIONS} 

We have solved the radiative transfer equation that describes
scattering of photons by cold electrons moving at relativistic speeds,
in steady-state, spherically symmetric accretion flows onto neutron
stars and black holes.  We used an iterative procedure to integrate
the radiative transfer equation along the photon characteristics.
This is an efficient method that allows for the full description of
the spacetime geometry and the velocity field (see Schmid--Burgk 1978
and Zane et al.\ 1996), independent of the optical thickness of the
flow. In this paper, we focused on examining simple configurations
with the intention of exploring the generation of hard
power-law spectra from accretion flows.  We have examined the
dependence of the resulting spectra on the boundary conditions, the
properties of the scattering medium, and the approximations often used
for the spacetime geometry and the velocity field.

We demonstrated that the generation of power-law spectral tails is the
result of multiple scatterings of the soft source photons by the fast
moving cold electrons (see also Laurent \& Titarchuk 1999), at both
low and high optical depths.  As a result, the photons emerging from
the accretion flow with high energies do not carry any signatures of
regions of high electron velocities or spacetime curvature -contrary
to expectations- but rather reflect the fact that a small number of
photons can experience a significant number of scatterings even in a
flow of small total optical depth.  This mechanism is very similar to
the generation of power-law spectra by thermal Comptonization in
static media. It differs, however, from the generation of power-law
spectra by non-thermal electrons in optically thin media, which is the
result of a single scattering by a power-law distribution of
relativistic electrons (see Coppi 1999).

The multiple scatterings experienced by the photons tend to wipe out
all memory of initial conditions such as the radial, angular, or
energy distribution of the sources. However, the resulting spectra
for {\it neutron-star} flows do depend sensitively on the choice of
the inner boundary conditions, confirming results reported earlier
(Titarchuk et al.\ 1997, Psaltis 1998; Psaltis \& Lamb 2000). It is
important to note, nevertheless, that none of the idealised boundary
conditions used in our calculations is, strictly speaking,
astrophysically relevant for a radial accretion flow onto a
neutron star. Indeed, whether the photons that reach the inner
boundary will be absorbed or reflected depends on the photon energy
and the ionization state of the surface layers of the neutron
star. Even if most of the photons are reflected, their energy gain
at reflection will be determined by the thermal state and
stratification of the transition region between the flow and the
stellar surface. Moreover, the presence of shocks at this interface
introduces additional complications. As a result, the spectral
signature of a nearly-radial accretion flow onto a neutron star will
not be uniquely determined by the flow properties but will be dictated
by the specifics of the interaction of the flow with the stellar
surface.

 On the contrary, the spectral indices of the hard spectral tails
resulting from the same flows onto black holes are immune to the inner
boundary condition. 
 This is due to the physical character of the 
inner boundary (the event horizon), i.e., that no radiation can escape 
to infinity (see
Fig.~\ref{fig:photon_escape} \& \ref{fig:ibc_effect}).  
For this reason, the spectra produced from bulk Comptonization 
in a {\em purely radial, free-falling} flow onto a black hole have a power law index that
is determined {\it solely} by the optical depth (or mass accretion rate
Eq.~[\ref{eq:m_dot}]; see also discussion in Titarchuk \ Zannias
1998).
Furthermore,
when compared to the spectral tails from flows onto neutron stars
in this idealised limit, those of  black holes are steeper.

Several previous studies have examined the generation of power-law
spectra without taking into account the curvature of the spacetime or
the regime of highly relativistic speeds (e.g., Payne \& Blandford
1981; Titarchuk et al.\ 1987; Psaltis \& Lamb 2000). We showed that,
for a free-fall velocity profile, general relativistic effects
identically cancel the bulk Comptonization effects {\em everywhere\/}
in the flow, including radii very far from the compact object, as long
as the photon mean-free path is very small.  If this were not true,
then an observer comoving with the flow would be able to distinguish
between being at rest at infinity and free-falling onto the compact
object by making only local (because of the small photon mean-free
path requirement) measurements of the evolution of the photon spectrum
with time. However, for typical accretion rates, nowhere in the flow
is the photon mean-free path small. Moreover, the photons that escape
to infinity have experienced their last scatterings in regions of
large photon mean-free path. As a result, the aforementioned
cancellation of ${\cal{O}}(\beta^2)$ terms reduces the Comptonization
efficiency --with respect to the flat geometry case-- only modestly in
the case of flows onto neutron stars which are of relatively high
optical depths.  Ignoring the actual spacetime geometry, when
modelling a flow onto a black hole, results in more severe
over-predicting of the hard tail flux, though.  For a black hole, the
existence of trapped characteristics (see \S\ref{sec:numerics})
deprives the outer parts of the flow from a large fraction of the
photons that have reached  close to the event horizon. Consequently,
disregarding the Schwarzschild spacetime geometry can lead to
overestimating the flux at high photon energies by more than an order
of magnitude (e.g., at 100~$T_{\rm in}$ in the case of
Fig.~\ref{fig:gr_flat_comparison}).

\mbox{} We thank Fred Lamb, Feryal \"Ozel, Philipppos Papadopoulos,
and Luca Zampieri for useful discussions and comments. This work was
supported by a postdoctoral fellowship of the Italian  MURST
(H.\,P.), a postdoctoral fellowship of the Smithsonian Institution and
also, in part, NASA (D.\,P.).  We also thank MPI f\"ur
Gravitationsphysik, Harvard-Smithsonian CfA, and University of
Portsmouth (H.\,P.) as well as SISSA (D.\,P.) for their warm
hospitality.

\end{document}